# Experimental realization of smectic phase in vortex matter induced by symmetric potentials arranged in two-fold symmetry arrays.

J. del Valle[1], A. Gomez[1], E. M. Gonzalez[1,2], M. R. Osorio[2], F. Galvez[1], D. Granados[2] and J. L. Vicent[1,2]

[1]Departamento Fisica de Materiales, Facultad de CC. Fisicas, Universidad Complutense, 28040 Madrid (Spain).
[2]IMDEA-Nanociencia, Cantoblanco, 28049 Madrid (Spain)

## Abstract

Smectic order has been generated in superconducting Nb films with two-fold symmetry arrays of symmetric pinning centers. Magnetic fields applied perpendicularly to the films develop a vortex matter smectic phase that is easily detected when the vortices commensurate with the pinning center array. The smectic phase can be turned on and off with external parameters.

## 1. Introduction

Crystals exhibit fully translational periodicity; on the other hand liquids do not show translational periodicity at all. In between smectic phase shows translation periodicity only in one dimension. Actually, smectic systems are solid-like in one direction and liquid-like in two directions. Liquid crystals are the paradigm of smectic phases; but we have to notice that these phases show up in a lot of systems besides liquid crystals. In addition, smectic order is claimed as the clue to understand many phenomena occurring in different systems. Examples, taken from very dissimilar systems, are nanorods with smectic order into patterned plasmonic nanostructures [1], smectic modulations in the pseudogap states of underdoped $Bi_2Sr_2CaCu_2O_{8+\delta}$ superconductor [2] and spontaneous ferroelectric order in a bent-core smectic liquid crystal [3]. Actually, the interest in smectic order is displayed beyond liquid crystal framework; for example de Gennes has explored the possible analogy between smectic phase and the mixed state phase in superconductors [4]. Carlson *et al.* examined the possibility of smectic phase in anisotropic superconductors [5] and Reichhardt *et al.* [6] showed that quenched disorder can induce a smectic phase. Among the systems which are good candidates to look for smectic phases, a promising choice is vortices in layered superconductors. We have to note that vortex matter is a very well established field; i.e. an ideal playground to test different models and going deeper in relevant features associated to phase transitions and related topic as is vortex lattice dynamics [7]. Concerning layered systems superconducting dichalcogenides [8], cuprates [9] and pnictides [10] have called the attention of many researchers. After a pioneer work [11] and some debate concerning the development of smectic order in vortex matter [12, 13], a smectic phase was experimentally found in cuprates [14], and in dichalcogenides [15].

In this paper, we show how to induce a vortex matter smectic phase in non-layered superconductors with periodic array of symmetric pinning potentials, and how this vortex matter phase can be easily handled.

## 2. Experimental Results

In layered superconductors the layers help to induce smectic order. In non-layered superconductors lack a suitable structure which can promote a smectic order. In layered materials, magnetic field applied parallel to the layers can trigger a smectic phase as has been reported in the literature [15, 16]. Layers allow placing and controlling the vortices easily. In the present work we dealt with plain superconductors (Nb films), so a different approach is needed. First of all, we need controlling vortices in plain superconductors. Arrays of non-superconducting centers embedded in the superconducting films are a right choice to accomplish this aim. Many researchers have studied vortices in superconducting films with artificially periodic pinning centers [17]. Superconducting films with periodic pinning nanocenter arrays show noteworthy effects which appear when matching between the vortex lattice and the array unit cell occurs. At matching conditions magnetoresistance shows minima. These minima show strong reduction of the dissipation and two neighbor minima are always separated by the same magnetic field value. The first minimum appears at magnetic field $H_1 = (\Phi_0/S)$, where S is the unit cell area of the pinning array and $\Phi_0 = 2.07 \; 10^{-15}$ Wb is the quantum fluxoid. Other minima appear at commensurability fields $H_n = n \; (\Phi_0/S)$, where n >1 is an integer number. Minima can be also observed at fractional matching fields $H_f = f \; (\Phi_0/S)$, being f a non-integer number. So, at matching conditions between the array and the vortex lattice, we are able to control the superconducting vortices. All the data presented in this work are taken with the magnetic field applied perpendicular to the film and with magnetic field values that fulfill the commensurability constraint, i. e. the applied field is a fraction or a multiple of the first matching field. We have to note that vortex behavior is governed by the interplay between random intrinsic pinning, which is known to be strong in Nb thin films [18], and artificially induced periodic potentials [19]. As explained by Pogosov et al. [20] at matching condition, both contributions are operational. Competition between these two pinning forces and elastic strains lead to the appearance of defects in the vortex lattice which break the long range translational symmetry, making the correlation length finite [21]. Therefore, a perfect ordered vortex lattice is absent.

In this work, the samples are 100 nm thick Nb films grown by sputtering on top of arrays of Cu dots (220 nm diameter and 40 nm thickness) which were fabricated on Si substrate by sputtering and electron beam lithography techniques. These nanodot dimensions yield a filling factor of one single vortex trapped for nanodot [22]. Finally, the samples are patterned in a cross-shaped bridge for magnetotransport measurements. More experimental details can be found in Ref. 22. To study the vortex phases in this type of samples we have followed the same approach that reported in Ref. 23-25; that is: i) the seminal paper of Fisher et al. [23] about glass to liquid second order phase transition and how to extract the critical exponents; ii) the work of Strachan et al. [24] regarding the carefully and unambiguously method to extract the critical temperatures; iii) the work of Villegas et al. [25] on periodic pinning and vortex glass phase, which shows that using a scaling analysis of I-V characteristics, they found that Nb thin films with periodic arrays of pinning centers show a continuous glass transition, similar to that observed in plain Nb films. The random and periodic pinning mechanisms compete and yield a glass phase which does not present long-range topological order.

We have grown a sample with an array of 400 x 400 nm$^2$ unit cell (SQ-sample in the following), that we use as standard sample for our study. We have measured (I,V) curves at several matching fields. The results do not depend of the value of the chosen matching field. Figure 1(a) shows I-V isotherms at H = 3H$_1$ (H$_1$ being the value of the first matching field H$_1$ = $\Phi_0$/a$^2$, with $\Phi_0$ = 2.07 10-15 Wb and a = 400 nm). As I →0, two different trends are observed. Isotherms close to T$_c$ show a linear dependence V $\propto$ I, this ohmic behavior corresponds to the vortex liquid phase. However, for lower temperatures it can be noted that I-V curves become highly non-linear for vanishing current, and voltage drops abruptly. This change corresponds to a transition to a non-dissipative vortex glass state. This melting transition is continuous, so we can define critical exponents ν and $z$ at which the phase correlation length of the glass $\xi_g$ ~ (T-T$_g$)$^{-\nu}$ and the relaxation time $\tau_g$ ~ $\xi_g^z$ diverge at the transition, T$_g$ being the melting temperature. Following Ref. 23, I-V data can be scaled down into two single curves according to:

$$\rho(1-T/T_g)^{\nu(D-2-z)} = f_{\pm}\{(J/T)(1-T/T_g)^{\nu(1-D)}\} \quad (1)$$

where D is the dimensionality of the system, ν and $z$ are the static and dynamic critical exponents respectively and f$_{\pm}$ are two scaling functions above and below T$_g$. Figure 1(b) shows scaling behavior for sample SQ at H = 3H$_1$, critical exponents ν = 1.0 ± 0.1 and $z$ = 6.7 ± 0.2 are obtained, in the range expected by the theory: ν ≈ 1-2 and $z$ ≈ 4-7. Dimensionality of the system is D=3.

In summary, we have found successful scaling analysis of the I-V data for sample SQ for magnetic fields which correspond to different matching fields obtaining ν and $z$ values in the ranges (1 ± 0.1, 1.1 ± 0.1) and (6.5 ± 0.2, 6.7 ± 0.2) respectively, supporting evidence of vortex glass to liquid transition in all cases, as expected.

Following Strachan *et al* [24], inset in Fig 1(b) shows the derivatives of log(V)-log(I) curves. We clearly observe the transition from ohmic to non-linear behavior at low currents. This crossover takes place at T$_g$, allowing us to determine the melting temperature from a direct and independent method, with an error of ±5 mK.

Once we have established the frame of our study the symmetry of the array is lowered from 4-fold to 2-fold symmetry. We have fabricated two samples one of them with array unit cell 400 x 600 nm$^2$ (R46 sample) and the another with array unit cell 400 x 800 nm$^2$ (R48 sample). The rectangular pinning landscape induces a strong anisotropy behavior in the vortex dynamics as was reported by Velez *et al* [26]. This anisotropic effect can be explored by (I,V) isotherm data taken with vortices moving along the short and the long sides of the rectangular unit cell. Several (I,V) curves were measured at different matching fields. Fig. 2 (a) and (b) show the (I,V) isotherm curves measured along the short and the long sides of the rectangular unit cell for the first matching field. The analysis of these raw data, following the same procedure than in SQ sample, leads to the following remarkable experimental facts in the low matching field regime: i) the experimental data cannot be scaled down, ii) using the log(V)-log(I) derivative analysis two different $T_g$ ($T_{gs}$, $T_{gl}$ in the following) are obtained. $T_{gs}$ is the transition temperature found for vortices moving along the short side of the array unit cell and $T_{gl}$ is the transition temperature obtained for vortices moving along the long side of the array unit cell.

This implies that for $T_{gs} < T < T_{gl}$, the low current behavior of the I-V curves is ohmic along one direction and non-linear along the other, so the system shows liquid or glass vortex dynamics depending on the direction.

3. **Discussion**

This behavior could be the hint to argue that in between these two temperatures ($T_{gs}$, $T_{gl}$) vortex matter behaves as smectic phase. The important feature of the smectic phase, which distinguishes from nematic phase, is that vortices are arranged in rows. This fact leads to two melting temperatures ruling out a nematic phase. The melting from glass to liquid only occurs along vortex motion parallel to the short side. Fig. 3(a) shows the potential landscape which helps to visualize the translation periodicity along the long side. This landscape is obtained taking into account that vortices move in the potential centers which are induced by the Cu nanodot arrays; i. e. the vortices have to probe the structure of the pinning array close to $T_c$. This vortex-nanodot interaction can be roughly estimated considering the volume of the vortex core within the nanodot volume, following Campbell and Evetts [27]. In this approach the coherence length $\xi$ plays the leading role in the interaction between the vortex core and the non-magnetic pinning centers (Cu nanodots in our case). The estimation of the vortex core is obtained from the Ginzburg-Landau coherence lengths which are extracted as usual from $H_{c2}$ (T) measurements. Figure 3 (a) shows a plot of this interaction potential when temperature is close to $T_c$ and therefore the coherence length $\xi$ is large. Interestingly, decreasing the temperature, so diminishing the coherence length, the overlapping potential, which mimics a layered structure, vanishes and the potential finally recaptures the pinning landscape induced by the array of nanodots, see Fig. 3 (b). Worth to point out that this temperature interval comprises the smectic region. These potentials are the background of the liquid-like behavior, while the periodicity of overlapping potentials supports the solid-like behavior in its perpendicular direction.

The difference between the two melting temperatures $T_{gs}$ and $T_{gl}$ is 60 mK at the first matching field, the low $T_{gs}$ being the value obtained when the vortices move along the short side of the array unit cell. This temperature difference diminishes increasing the applied magnetic fields. Scaling down the experimental data is only possible when the two $T_g$ merge and a vortex glass to vortex liquid transition is recovered with usual values of the critical exponents, for example the critical exponents for H= $5H_1$ are ν =1.0 ± 0.1, and $z$ = 6.6 ± 0.2. See Fig 3(c) for a complete phase diagram picture.

We have also measured (I,V) curves for several matching fields in sample R48; in which the long side of the array unit cell (800 nm) is doubled in comparison with the short side (400 nm). Fig. 4 shows the (I,V) raw data for the first matching field. Alike sample 46, for the first matching field ($H_1$) in sample R48, the experimental data cannot be scaled down and two $T_g$ are obtained. In this case the temperature difference is 250 mK, more than four times the value found in sample R46 (60 mK). Scaling down the experimental data is only possible when the two $T_g$ merge, but in sample R48 the critical exponents depend on the vortex motion direction in the whole range that we have measured, for instance for $17H_1$, the critical exponents extracted from the scaling for vortex motion along the short or long sides of the array unit cell are: $ν_{short}$ = 1.1 ± 0.1, $z_{short}$ = 7.0 ± 0.2 and $ν_{long}$=1.1 ± 0.1, $z_{long}$ = 5.5 ± 0.2 respectively. Therefore, we are dealing with an anisotropic vortex glass to vortex liquid melting transition [28] in the high applied magnetic field region.

We can clarify this complex picture by studying how resistivity changes in a transition from the vortex liquid into a vortex smectic or glass phase (see Ref. 14, 28 and 29). In both cases, close to the transition the resistivity drops to zero as a power law, i.e.

$$\rho \sim (T/T_g - 1)^{s} \qquad (2)$$

Critical exponent **s** can be obtained in a direct way: dividing ρ by its derivative δρ/δT and finding the slope of the resulting curve (see the inset of Fig.5 for sample R46). Particularly, for a vortex liquid to glass transition Eq. (2) can be derived from Eq.(1), and this critical exponent will be **s** = ν(z+2-D) [30]. In that case the expected **s** values could be 3 ≤ **s** ≤ 12, taking into account D = 3 and the limits for both exponents (ν ≈ 1-2 and $z$ ≈ 4-7). The control sample SQ was measured and analyzed for several matching fields. In Fig. 5 we can see that the extracted **s** values are between 5 and 6, in the range expected for a vortex liquid to glass transition, and they do not depend on the magnetic field.

In the case of the 2-fold symmetry samples (R46 and R48), resistivity also follows Eq. (2) (see inset in figure 5) and the critical exponent **s** can be estimated. Figure 5 shows the results for both samples with selected applied magnetic fields which are fractional and multiple of the first matching fields. The **s** exponents show values that depend on the matching fields and they are lower than in the case of the control sample (SQ sample). We can notice two regimes: i) The **s** exponents are lower than 3 for matching fields lower than $H_3$. These low values of the **s** exponents have been reported previously for smectic phases [4, 14, 29, 31]. ii) For matching fields higher than n= 3, the critical exponent **s** rises, reaching, at the end, values that are between the expected values in the usual vortex glass-vortex liquid transition. Below the crossover (low applied magnetic field regime) both sample behavior looks similar, above the crossover (high applied magnetic field regime) the samples follow a different behavior. Sample R48 shows lower values and flatter behavior than sample R46. We have to remember that sample R48 shows an anisotropic scaling. In this case, the exponent values are in the lower limit of the expected values for the solid to liquid transition; i.e. the transition is smoother than usual, the resistivity within the vortex liquid state drops to zero as a power law with lower exponents; i. e. less abrupt transition than the isotropic transition.

Finally, concerning the crossover, we have to address two experimental facts: i) the smectic phase is only observed for small applied magnetic fields and ii) this happens around the same magnetic field (around $3H_1$) in both samples. The interplay between potential landscapes and vortices could be a hint to figure out these two findings. First of all, an increase of the number of vortices in the array unit cell smears out the vortex-nanocenter interaction, for example matching field of $3H_1$ means one trapped vortex and two interstitial vortices per unit cell, hence the vortex lattice - pinning potential interaction is weaker than in the case of $H_1$ (only trapped vortices) and $2H_1$ (trapped vortices and only one interstitial vortex). In conclusion, the weakness of the vortex-pinning landscape interaction precludes the smectic phase and the liquid phase is promoted. On the other hand a comparison between sample R48 and R46 shows that the translation periodicity along the long side is distinct in each one of the samples, but in the perpendicular direction the same potential landscape (400 nm between Cu dots) is found for both samples. Therefore, both samples look alike from this point of view. This could be the clue for finding a crossover at similar matching fields, since vortices probe the same potential landscape when they move along the short side of the rectangular unit cell.

## 4. Conclusions

In summary, plain superconducting Nb films can show a (H,T) phase diagram with a smectic region between the liquid and the solid phases. This is realized when the films are grown on top of array of symmetric pinning centers. Interestingly, smectic order is achieved when the symmetry of the array is reduced from 4-fold to 2-fold. That is, in these non-layered superconductors, vortex matter shows a liquid-like or solid-like behavior depending on the vortex motion direction. This potential landscape is fabricated with a two-fold symmetric array of Cu nanodots embedded in the superconductor. The smectic phase is controlled by the array shape, temperature and applied magnetic field. Finally, this smectic phase always vanishes increasing the number of vortices and the usual vortex phase diagram is recovered with vortex glass to vortex liquid crossover.

We thank Spanish MINECO grant FIS2013-45469 and CM grant S2013/MIT-2850 and EU COST Action MP-1201. D.G. acknowledges RYC-2012-09864.

Figure Captions

FIG. 1  (a) I-V isotherms from 0.969 $T_c$ to 0.997 $T_c$, data taken every 10 mK, for sample SQ ($T_c$ = 8.0 K) and H=380 Oe (H = $3H_1$). Isotherms above $T_g$ open red circles and below $T_g$ blue dots. (b) Scaling of the I-V data into two curves corresponding to the vortex liquid (red hollow dots) and glass (blue dots) phases. Derivatives of the log (V) - log (I) curves as a function of the current are plotted in the inset. $T_g$ = 0.989 $T_c$, as well as critical exponents ν = 1.0 and z = 6.7 are obtained (see text). (c) Sketch of sample SQ: array of Cu dots with 400 x 400 $nm^2$ unit cell embedded in Nb film grown on Si substrate. (Sketch is not to scale).

FIG. 2  I-V isotherms for sample R46 ($T_c$ = 8.7 K), from 0.978 $T_c$ to 0.995 $T_c$, data taken every 10 mK with H=86 Oe (H=$H_1$) for vortices moving (a) along the short and (b) along the long sides of the rectangular unit cell respectively. Depending on the direction, two different $T_g$ (solid lines) are obtained (see text): $T_{g,s}$ = 0.983 $T_c$ and $T_{g,l}$ = 0.990 $T_c$, with a 60 mK difference between both temperatures (green hollow squares) which spans between the vortex glass (blue dots) and vortex liquid (red dots). (c) Sketch of sample R46: array of Cu dots with 400 x 600 $nm^2$ unit cell embedded in Nb film grown on Si substrate. (Sketch is not to scale).

FIG. 3 (a) Pinning potential (U) generated by sample R46 for ξ = 95 nm corresponding to 0.988 $T_c$. ξ being the coherence length. (b) Pinning potential (U) generated by sample R46 for ξ = 70 nm corresponding to 0.980 $T_c$. (c) Phase diagram (T, H) Y-axis: $T_c$ (red squares), $T_{g,l}$ (green triangles) and $T_{g,s}$ (blue dots). X-axis $H/H_1$ in log scale, $H_1$ being the first matching field. VL (vortex liquid), VS (vortex smectic) and VG (vortex glass) see text.

FIG. 4 I-V isotherms for sample R48 ($T_c$ = 8.3 K) data taken every 10 mK with H=63 Oe (H=$H_1$) : (a) from 0.988 $T_c$ to $T_c$, vortices move along the long side of the rectangular array; (b) from 0.959$T_c$ to $T_c$ vortices move along the short side of the rectangular array. $T_{gl}$ =0.997 $T_c$ and $T_{gs}$ = 0.967 $T_c$. Green hollow squares show the experimental data which spans between the vortex glass (blue dots) and vortex liquid (red dots). (c) Sketch of sample R48: array of Cu dots with 400 x 800 $nm^2$ unit cell embedded in Nb film grown on Si substrate. (Sketch is not to scale).

FIG. 5 Y- axis critical exponent **s**, X-axis $H/H_1$ in log scale, $H_1$ being the first matching field for all samples. Sample SQ (black squares) and samples R46 (blue circles) and R48 (red triangles). The lines are guides to the eye. In the inset, linear fit to obtain **s**, as an example in sample 46. Resistivity drops as $(T-T_g)^s$.

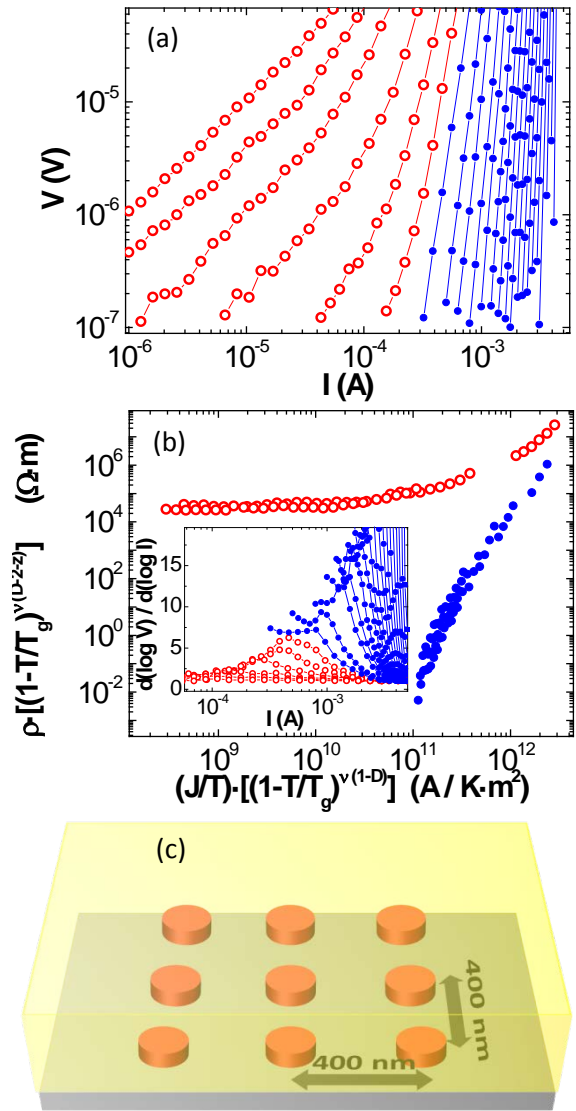

Fig. 1

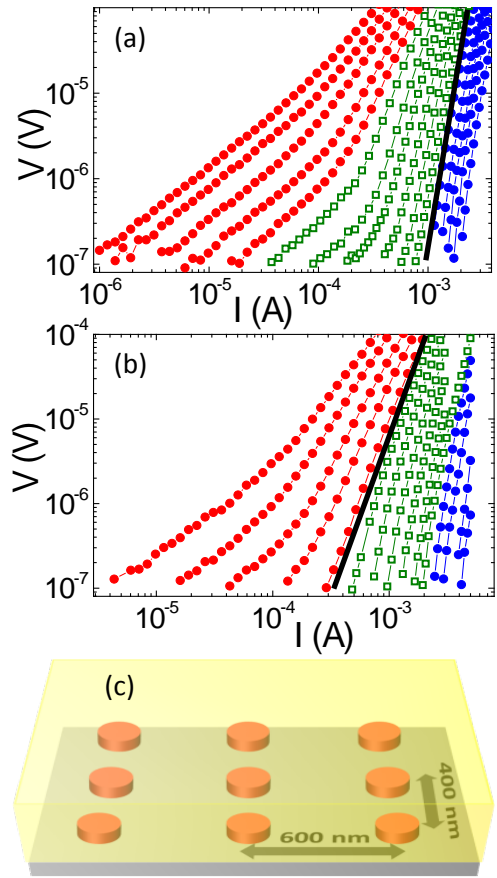

Fig. 2

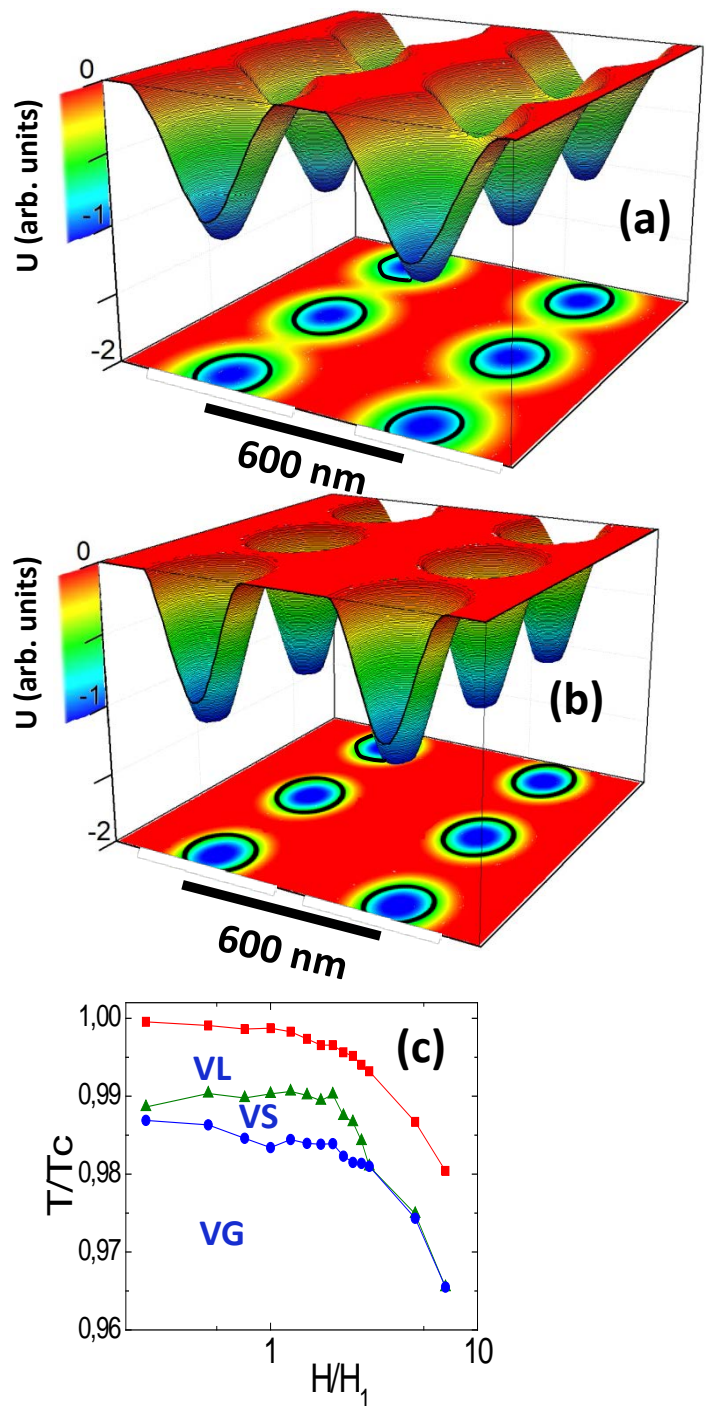

Fig. 3

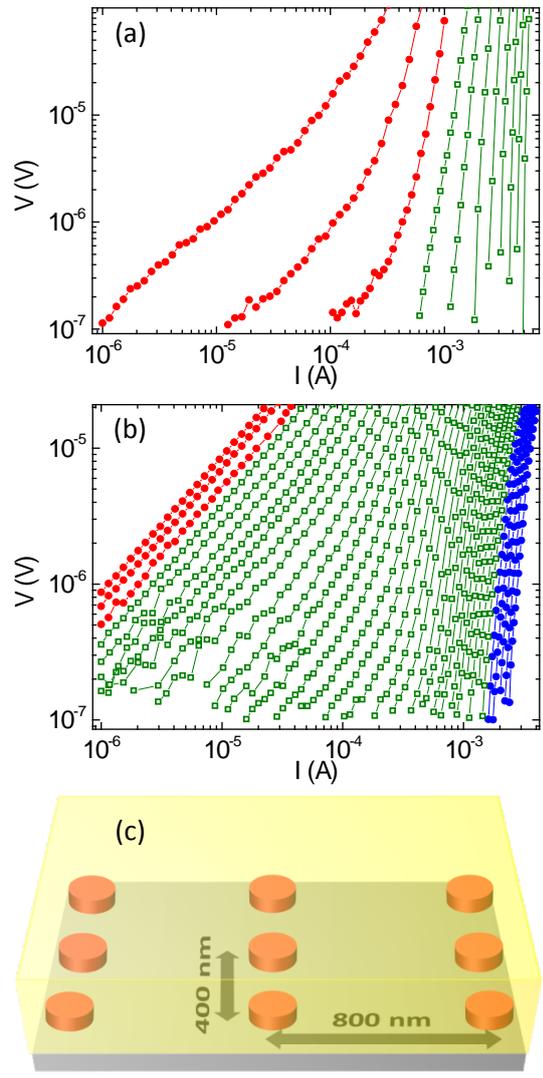

Fig. 4

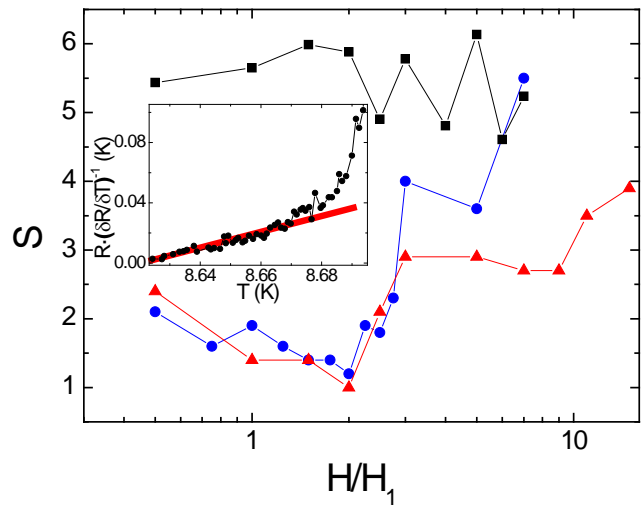

Fig. 5